
\input phyzzx
\overfullrule=0pt
\tolerance=5000
\twelvepoint
\def\npb#1#2#3{{\it Nucl.\ Phys.} {\bf B#1} (19#2) #3}
\def\plb#1#2#3{{\it Phys.\ Lett.} {\bf #1B} (19#2) #3}
\def\pla#1#2#3{{\it Phys.\ Lett.} {\bf #1A} (19#2) #3}
\def\prl#1#2#3{{\it Phys.\ Rev.\ Lett.} {\bf #1} (19#2) #3}
\def\mpl#1#2#3{{\it Mod.\ Phys.\ Lett.} {\bf A#1} (19#2) #3}
\def\prd#1#2#3{{\it Phys.\ Rev.} {\bf D#1} (19#2) #3}
\def\pr0#1#2#3{{\it Phys.\ Rev.} {\bf #1} (19#2) #3}
\def\p{\partial}
\def\coset{SL(2,{\bf R})/U(1)}
\def\ztwo{{\bf Z}_2}
\def\sltwo{SL(2,{\bf R})}
\def\soto{SO(2,1)}
\def\sooo{SO(1,1)}
\def\CP{{\cal P}}
\def\CM{{\cal M}}
\def\CT{{\cal T}}
\def\frac#1#2{{#1 \over #2}}

\def\sfrac#1#2{\hbox{${#1 \over #2}$}}
\def\CL{{\cal L}}
\def\CB{{\cal B}}
\def\semi{\hbox{${\;\subset\!\!\!\!\!\!\times\;}$}}

\def\bra#1{\langle #1 |}
\def\ket#1{|#1\rangle}

\def\endli{\hfill\break}
%
\pubnum{EFI-92-10}
\date{February 1992}
\titlepage
\title{\centerline{Two Dimensional Stringy Black Holes}
\centerline{with One Asymptotically Flat Domain}}
%
\vglue-.25in
\author{Petr Ho\v{r}ava\foot{e-mail addresses:
horava@yukawa.uchicago.edu or horava@curie.uchicago.edu}%
\foot{Robert R. McCormick Fellow; research also supported by the NSF under
Grant No.\ PHY 90-00386; the DOE under Grant No.\ DEFG02-90ER40560;
the Czechoslovak Chart 77 Foundation; and the \v{C}SAV under Grant No.\
91-11045.}}
\medskip
\address{\centerline{Enrico Fermi Institute}
\centerline{University of Chicago}
\centerline{5640 South Ellis Avenue}
\centerline{Chicago, IL 60637, USA}}
\bigskip
\abstract{The exact black hole solution of 2D closed string theory has, as any
other maximally extended Schwarzschild-like geometry, two asymptotically flat
spacetime domains.  One can get rid of the second domain by gauging the
discrete symmetry on the $\sltwo /U(1)$ coset that interchanges the two
asymptotic domains and preserves the Kruskal time orientation everywhere in
the Kruskal plane.  Here it is shown that upon performing this orbifold
procedure, we obtain a theory of unoriented open and closed strings in a black
hole background, with just one asymptotically flat domain and a time-like
orbifold singularity at the origin.  All of the open string states of the
model are confined to the orbifold singularity.  We also discuss various
physical aspects of the truncated black hole, in particular its target
duality -- the model is dual to a conventional open string theory in the
black hole geometry.}
\endpage
%
%
\REF\witten{E. Witten, \prd{44}{91}{314}}
\REF\MSW{G. Mandal, A.M. Sengupta and S.R. Wadia, \mpl{6}{91}{1685}}
\REF\ancient{I. Bars and D. Nemeshansky, \npb{348}{90}{89}\endli
S. Elitzur, A. Forge and E. Rabinovici, \npb{359}{91}{581}\endli
M. Ro\v{c}ek, K. Schoutens and A. Sevrin, \plb{265}{91}{303}}
\REF\dvv{R. Dijkgraaf, E. Verlinde and H. Verlinde, Princeton preprint
IASSNS-HEP-91/22 = PUPT-1252 (May 1991)}
\REF\marsh{E.J. Martinec and S.L. Shatashvili, \npb{368}{92}{338}}
\REF\bars{I. Bars, USC preprints USC-91/HEP-B3 (May 1991), USC-91/HEP-B4
(June 1991)}
\REF\ellis{J. Ellis, N.E. Mavromatos and D.V. Nanopoulos, \plb{267}{91}{465};
{\bf 272B} (1991) 261; CERN/Texas A\&M preprints CERN-TH.6309/91 =
CTP-TAMU-90/91 (November 1991) and CERN-TH.6351/91 = CTP-TAMU-100/91
(December 1991)}
\REF\giveon{A. Giveon, \mpl{6}{91}{2843}}
\REF\kiritsis{E.B. Kiritsis, \mpl{6}{91}{2871}}
\REF\rocek{M. Ro\v{c}ek and E. Verlinde, IAS/Stony Brook preprint
IASSNS-HEP-91/68 = ITP-SB-91-53 (October 1991); A. Giveon and M. Ro\v{c}ek,
IAS/Stony Brook preprint IASSNS-HEP-91/84 = ITP-SB-91-67 (December 1991)}
\REF\nelson{J. Distler and P. Nelson, Princeton/Penn preprint
PUPT-1262 = UPR-0462T (August 1991)}
\REF\beku{M. Bershadsky and D. Kutasov, \plb{266}{91}{345}}
\REF\sixties{W. Rindler, \prl{15}{65}{1001};  W. Israel, \pr0{143}{66}{1016};
J.L. Anderson and R. Gautreau, \pla{20}{66}{24};  F.J. Belinfante,
\pla{20}{66}{25}}
\REF\gibb{G.W. Gibbons, \npb{271}{86}{497}, and Carg\`{e}se lectures, 1985}
\REF\bershkut{M. Bershadsky and D. Kutasov, \plb{274}{92}{331}}
\REF\openbh{M.D. McGuigan, C.R. Nappi and S.A. Yost, IAS preprint
IASSNS-HEP-91/57 (October 1991)}
\REF\wso{P. Ho\v{r}ava, \npb{327}{89}{461}}
\REF\etsm{P. Ho\v{r}ava, Prague preprint PRA-HEP-90/18 (December 1990)}
\REF\toptorus{P. Ho\v{r}ava, Chicago preprint EFI-92-06 (January 1992), and
to appear}
\REF\shenker{S. Shenker, Rutgers preprint RU-90/47 (August 1990),
and private communication}
\REF\vilen{N.Ya.\ Vilenkin, ``Special Functions and the Theory of Group
Representations'' (Moscow, Nauka, 1965)}
\REF\haha{J.B. Hartle and S.W. Hawking, \prd{13}{76}{2188}}
\REF\dual{P. Ho\v{r}ava, \plb{231}{89}{351}}
\REF\boundcross{C.G. Callan, C. Lovelace, C.R. Nappi and S.A. Yost,
\npb{293}{87}{83}\endli J. Polchinski and Y. Cai, \npb{296}{88}{91}}
\REF\csworb{P. Ho\v{r}ava, Prague preprint PRA-HEP-90/3 (June 1990)}
\REF\atick{J.J. Atick and E. Witten, \npb{310}{88}{291}}
\REF\dlp{J. Dai, R.G. Leigh and J. Polchinski, \mpl{4}{89}{2073}}
\REF\membr{J. Bagger, C.G. Callan and J.A. Harvey, \npb{278}{86}{550}\endli
C.G. Callan, J.A. Harvey and A. Strominger, \npb{359}{91}{611}}
\REF\ucsb{D. Garfinkle, G.T. Horowitz and A. Strominger, \prd{43}{90}{3140};
N. Ishibashi, M. Li and A.R. Steif, Santa Barbara preprint UCSBTH-91-28
(June 1991)}
\REF\eguchi{T. Eguchi, \mpl{7}{92}{85}}
\REF\polch{J. Polchinski, \npb{362}{91}{125}\endli D. Minic, J. Polchinski
and Z. Yang, \npb{369}{92}{324}}
\REF\moore{G. Moore, \npb{368}{92}{557}; G. Moore, M.R. Plesser and S.
Ramgoolam, Yale preprint YCTP-P35-91 (November 1991)}
%
%
Black holes are spacetimes of amazingly rich geometrical and topological
structure.  The recent discovery of exact black hole solutions of 2D string
theory in terms of an $\sltwo /U(1)$ coset CFT ([\witten,\MSW], see also
[\ancient]) allows one to study some properties of quantum theory in the
geometry of a black hole quite explicitly [\dvv,\marsh,\bars,\ellis], and in
particular to search for stringy effects, such as duality
[\giveon,\kiritsis,\rocek].  Another important aspect of the black hole
solution is its expected relation to $c=1$ matter coupled to Liouville
[\witten,\marsh,\nelson,\beku].

One of the many interesting issues of quantum black hole physics is the role
of various asymptotically flat domains in the black hole geometry.  As any
other maximally extended Schwarzschild-like black hole, the 2D black hole
has two distinct asymptotically flat exterior regions (denoted by I and II
throughout the paper) which cannot influence each other causally.
In this paper we will address the question of getting rid of the second
asymptotically flat domain of the stringy black hole in a consistent way.%
\foot{Considerable effort in this direction has been made in the 4D black hole
physics since the middle of sixties [\sixties] (for a review, see e.g.\
[\gibb]).  Most of the authors have identified points $(u,v)\equiv (-u,-v)$ in
the Kruskal plane, which is not the strategy accepted in this paper.}
Actually, we will point out that the truncated black hole with just one
asymptotically flat exterior domain is naturally inhabited by unoriented
open and closed strings.  As any difference between closed and open strings
gets lost in the point particle limit, this is a typical stringy effect
in the black hole physics.  After establishing this result, we will analyze
more closely some aspects of the truncated black hole, in particular its
spectrum, spacetime duality, and the description of the black hole
singularity in terms of an effective topological field theory.  (Some other
aspects of black holes in 2D open string theory have been studied recently
in [\bershkut,\openbh].)

In string theory, there is a particularly natural way of truncating spacetime
manifolds, by using orbifold techniques.  In fact, this strategy has been
recommended for the 2D black holes quite explicitly by Witten [\witten], who
has proposed to mod out the black hole coset by the $\ztwo$ symmetry that
maps the two asymptotically flat domains to each other by mapping $u$ to
$v$ in the Kruskal-Szekeres coordinate system.  Once we try to do so
throughout the Kruskal plane, we observe that we have to supplement
the $\ztwo$ action on the target by an orientation reversal on the
worldsheet.  Orbifolds of this type are known to lead to the theory of open
and closed unoriented strings [\wso].  In other words, the proposed orbifold
model will represent a typical example of the worldsheet orbifold
construction in the sense of [\wso], hence leading to a target with both
closed and open strings in the spectrum.

This situation resembles somewhat the theory of topological sigma models with
K\"{a}hler targets [\etsm]: Once one attempts to mod out the topological
target by an antiholomorphic involution, one has to act simultaneously by a
complex conjugation on the worldsheet.  Such a combined $\ztwo$ action can be
gauged, and the resulting theory describes topological sigma models of open
string theory (or $\ztwo$-equivariant topological sigma models in the
terminology of [\etsm]).  This analogy is perhaps not accidental, as there
are some indications [\toptorus] that two dimensional string theory, of which
the black hole is a classical solution, might have an underlying topological
phase.

Before actually discussing our truncated black hole, let us mention some more
motivation.  First, the role of various spacetime geometries in 2D string
theory should be probably addressed in a non-perturbative framework, such as
string field theory.  Up to now, string field theory of open strings is still
conceptually much better understood than its closed string counterpart.
Second motivation comes from the fact that, as has been argued some time ago
by Shenker [\shenker], the leading nonperturbative effects in matrix models
are of order $e^{-C/\kappa}$, where $\kappa^2$ is the closed string coupling
constant.  This is a stringy effect, as in quantum field theory one rather
expects the leading nonperturbative effects to behave like $e^{-C/\kappa^2}$.
Recalling that $\kappa$ is the open string coupling constant, it is tempting
to speculate that open strings are hidden somehow inside the closed string
theory, and generate the leading nonperturbative effects of order
$e^{-C/\kappa}$.  It is thus natural to search for a mechanism in which open
strings naturally emerge in low dimensional closed string theories.  We hope
to offer one in this paper.

To answer the physical question of whether one can forget consistently about
the second asymptotically flat domain of the stringy black hole, we will
construct the truncated black hole by gauging a $\ztwo$ extension of $U(1)$
(or of its non-compact version $\sooo$) in the black hole coset.  First,
after summarizing some basic facts to fix our notation, we will identify the
corresponding $\ztwo$ action.

The Lagrangian for the $\sltwo$ WZW model
$$\CL_{\rm WZW} = {k \over 8\pi}\int _\Sigma d^2z\;
{\rm Tr}\; (g^{-1}\p g\; g^{-1}\bar{\p} g)+{ik \over
12\pi}\int _B{\rm Tr}\; (g^{-1}dg\wedge g^{-1}dg\wedge g^{-1}dg)\eqn\lagwzw$$
is by construction symmetric under the left-moving and right-moving current
algebras of $\sltwo$, acting on an element $g$ of $\sltwo$ by
$$g(z,\bar{z}) \rightarrow \Omega_L(z)\,  g(z,\bar{z}) \,\Omega_R(\bar{z}).
\eqn\actloc$$
For the study of the maximally extended Minkowski black hole in Schwarzschild
coordinates, the following faithful parametrization [\vilen] of $\sltwo$ is
most suitable:
$$g=\pmatrix{e^{t_L/2}&0\cr 0&e^{-t_L/2}\cr} \pmatrix{-
1&0\cr 0&-1\cr}^{\epsilon_1}\pmatrix{0&-1\cr 1&0\cr}^{\epsilon_2}
\CM \pmatrix{e^{-t_R/2}&0\cr 0&e^{t_R/2}\cr},\eqn\param$$
where $t_L ,t_R \in {\bf R}$, $\epsilon_1, \epsilon_2 \in\{ 0,1\}$ and $\CM$
is one of the following two matrices:
$$\CM =\left\{ \eqalign{
\pmatrix{\cosh\frac{r}{2}&\sinh\frac{r}{2}\cr
\sinh\frac{r}{2}&\cosh\frac{r}{2}\cr} & \qquad\qquad
r\in {\bf R},\cr
\pmatrix{\cos\frac{r}{2}&-\sin\frac{r}{2}\cr\sin\frac{r}{2}&
\cos\frac{r}{2}\cr}&\qquad\qquad
r\in (-\pi /2,\pi /2).\cr}
\right. \eqn\pmat$$
$\epsilon_1$ parametrizes the centrum of $\sltwo$, and is irrelevant on
$\soto$, while $\epsilon_2$ is related to the (self) duality of the black
hole.

We can gauge any non-anomalous subgroup of the symmetry group.  The
requirement of non-anomalousness leaves just two choices for gauging Abelian
symmetries [\kiritsis], namely the vector symmetry $g\rightarrow hgh^{-1}$
and the axial symmetry $g\rightarrow hgh$.  Upon gauging the axial $U(1)$
generated by $\sigma _3={\rm diag}(1,-1)$ and choosing the unitary gauge,
$$t_L=t_R\equiv t,\eqn\timedef$$
the first choice of $\CM$ as indicated in \pmat\ describes regions I and II
(and regions V and VI behind the singularities, depending on the value of
$\epsilon_2$), while the second choice of $\CM$ describes the interior of the
black hole between the horizon and the singularity (regions III and IV).  The
conventional Kruskal coordinates, $u$ and $v$, are related to the
Schwarzschild coordinates $r,t$ by
$$g=\pmatrix{a&u\cr -v&b\cr}, \qquad ab+uv=1,\eqn\krusk$$
and the unitary gauge \timedef\ is now $a=\pm b$.

The Euclidean black hole geometry, on the other hand, is most transparent in
the Euler angle parametrization of $SU(1,1)$ (isomorphic to $\sltwo)$:
$$h=\pmatrix{\alpha&\beta\cr\bar{\beta}&
\bar{\alpha}\cr}=\pmatrix{\cosh\frac{r}{2}\; e^{\frac{i}{2}(\theta_R-\theta_L
)}&\sinh\frac{r}{2}\; e^{\frac{i}{2} (\theta_L+\theta_R )}
\cr \sinh\frac{r}{2}\; e^{-\frac{i}{2}(\theta_L+\theta_R )}
&\cosh\frac{r}{2}\; e^{\frac{i}{2}(\theta_L-\theta_R )}\cr},
\eqn\suoneone$$
with
$$\theta_L\in {[}0,2\pi ), \qquad r\in [0,\infty )
,\qquad \theta_R \in [-2\pi ,2\pi ).\eqn\slpar$$
Note that we can obtain \suoneone\ from \param\ by an analytic continuation,
setting $t_{L,R}=i\theta_{L,R}$.  In the unitary gauge, this is the Euclidean
continuation of the black hole, in which the Schwarzschild time coordinate is
continued analytically.  Later on we will be using another Euclidean
continuation of the Minkowski black hole, used e.g.\ by Hartle and Hawking in
[\haha], in which the coordinate to be continued is the Kruskal time.

At least two discrete symmetries of the coset are worth mentioning.  First,
the theory is symmetric under the $\ztwo$ that multiplies $g$ by the
non-trivial central element of $\sltwo$.  What is usually done in this
situation is the obvious modification of the WZW model to the orbifold model
based on the centerless factor of $\sltwo$, isomorphic to $\soto$.  This
procedure is profitable because it eventually takes away the two-sheeted
degeneration of the Minkowski black hole geometry mentioned first in
[\witten].  We will interpret the black hole geometry in precisely this sense.
\foot{The two-sheeted geometry of the $\sltwo /\sooo$ black hole can be
interpreted alternatively as follows.  Generic points in the Penrose diagram
of a $D$-dimensional black hole describe $(D-2)$-spheres, isomorphic in two
dimensions to the set of two elements, $\{ 1,-1\}$.  In this sense, the black
hole will have four asymptotically flat domains, two of them describing the
two sides of one ``universe'' surrounding the black hole, and the remaining
two describing the ``mirror universe.''  We will not accept this
interpretation in this paper, however.}
Another important $\ztwo$ symmetry of the coset model is generated by the
inversion of $g$,
$$\Theta ' :g\rightarrow g^{-1}.\eqn\inve$$
This is an anti-automorphism of the group, hence it sends $\Omega_{L,R}$ to
$\Omega_{R,L}^{-1}$ in \actloc .  As $\Omega_L$ is purely left moving and
$\Omega_R$ purely right moving, $\Theta '$ has to acs simultaneously as a
parity reversal on the worldsheet in order to become a symmetry of the coset.

Now we are going to find the orbifold action that will map one asymptotically
flat region to the other by mapping $u$ to $v$ and vice versa.  To lowest
order in $1/k$, the spacetime metric and dilaton field are given by
$$\ \ \ \ \ ds^2=\frac{du\, dv}{1-uv}, \qquad \Phi=\ln (1-uv)+{\rm const}.
\eqn\lowestmetric$$
This geometry solves the low-energy effective approximation to classical
string theory in two dimensions.  Clearly, the intended orbifold action is a
symmetry of \lowestmetric ;  the crucial point is to extend the symmetry to
the full-fledged classical string solution, represented by the exact coset.

The corresponding $\ztwo$ action can be identified e.g.\ by requiring the
correspondence with the flat limit of the black hole geometry, where the
model reduces to free fields, or (with some guesswork) even directly;
the action we need is given by
$$\Theta :\quad g\;\rightarrow\;\pmatrix{0&1\cr 1&0\cr}g^{-1}
\pmatrix{0&1\cr 1&0\cr}\equiv sg^{-1}s.\eqn\actwso$$
To be a symmetry of the coset, $\Theta$ has to act simultaneously on the
worldsheet as the parity flip.  $\Theta$ extends the Abelian $\sooo$ gauge
group to the direct product $\ztwo\times\sooo$.  As the gauge group now acts
on the worldsheet as well as in the target, we have arrived at a typical
example of orbifold models studied in [\wso], and expect open strings to
emerge in the twisted sector of the model.  As we will see below, this is
indeed true, and establishes the central result of the paper.

We have just identified our orbifold group action, leading to a worldsheet
orbifold model. One can note, however, that the coset enjoys also a purely
{\it target} $\ztwo$ symmetry that maps region I to region II by
$u\leftrightarrow v$, or in  terms of the group variable, acts by
$$g\rightarrow \pmatrix{0&-1\cr 1&0\cr}g\pmatrix{0&1\cr -1&0\cr}.\eqn\tarorb$$
This discrete symmetry can be gauged, leading to a target orbifold model
with one asymptotic region, orbifold singularity at $u=v$ in regions III and
IV, and just closed strings in the spectrum.  Does the existence of this
orbifold model endanger our claim that the truncated black hole is inevitably
a theory of open strings?

The answer to this question results from a careful analysis of the orbifold
action on the full-fledged coset.  In the Kruskal parametrization, \tarorb\
becomes
$$\pmatrix{a&u\cr -v&b\cr}\rightarrow \pmatrix{b&v\cr
-u&a\cr}.\eqn\tarorbkr$$
In the unitary gauge we set $a=b$ in regions I$-$IV and $a=-b$ in regions
V, VI.\ \  Recalling now that two elements of $\sltwo$ that differ by an
overall minus sign describe actually the same point on $\soto$, we define
Kruskal coordinates $u,v$ on the one-sheeted black hole geometry by requiring
non-negative $a$ in \tarorbkr .  (This amounts to a gauge fixing of the
discrete gauge symmetry that makes the $\soto$ WZW model out of the $\sltwo$
WZW model.)  Keeping in mind these gauge fixing conditions, one can see that
the target orbifold action \tarorbkr\ does indeed map $u$ to $v$ in regions
I$-$IV, but interchanges regions V and VI (transforming $(u,v)$ to $(-u,-v)$
there).  This target orbifold might represent an attractive possibility for
getting rid of the second asymptotic region, as the action \tarorb\ on the
target is everywhere space-like.  Nevertheless, the worldsheet orbifold
\actwso\ remains the only possibility for modding out by $u\leftrightarrow v$
everywhere, and will be the only orbifold analyzed in this paper.

Let us now analyze the spectrum of the truncated black hole, first in the
Euclidean signature.%
\foot{Since our orbifold action mixes different causal
regions of the Minkowski black hole, it is crucial -- in order to keep the
physical interpretation of the orbifold unchanged -- to use the Hartle-Hawking
Euclidean continuation [\haha], which covers the whole Kruskal plane.}
 Physical states of the black hole are defined as cohomology classes of the
BRST charge for the coupled system of the coset matter and diffeomorphism
ghosts on the worldsheet,
$$Q=Q_{\rm coset}+Q_{\rm gravity},\qquad Q_{\rm coset}=
\oint_Cc(J^3+ikA)+{\rm h.c.}\eqn\brstcharge$$
(here $c$ is the spin-zero $U(1)$ ghost, and $A$ is the $U(1)$ gauge field).
The orbifold group action defines its own condition of physicality, as it
projects the physical states of the full black hole to the $\ztwo$ invariant
sector (and introduces twisted states).  This can be reformulated in terms of
equivariant cohomology of the BRST charge, thus defining the physical states
of the model as the cohomology of $Q$, equivariant with respect to the
antighost field action [\nelson] as well as the orbifold $\ztwo$ action.
In the closed string sector, these conditions are
$$Q|{\rm phys}\rangle =0,\qquad
(b_0-\bar b_0)|{\rm phys}\rangle =0,\qquad
(\Theta -1)|{\rm phys}\rangle =0.\eqn\cohomolo$$
The model to be orbifoldized is the $\sltwo /U(1)$ coset, which is a target
for closed string propagation.  Hence, the untwisted sector on the orbifold
consists of the closed strings of the original model.  The only thing that
remains to be identified in the untwisted sector is the action of the
orbifold group on it.  It is illuminating to weaken \cohomolo\ temporarily,
and study primaries of the coset.  Parametrizing the group by \suoneone ,
vertex operators of the $\coset$ primaries of the closed-string black hole
CFT are given by matrix elements of (unitary) $\sltwo$ representations:
$$T^{\ell}_{mn}(r,\theta_L,\theta_R)=\left\langle \ell
,m\right| g(r,\theta_L,\theta_R)\left|\ell ,n\right\rangle .\eqn\vert$$
This expression can be written with the use of Jacobi functions as
$$T^{\ell}_{mn}(\theta_L,r,\theta_R)=\CP^{\ell}_{mn}
(\cosh r)e^{im\theta_L+in\theta_R}.\eqn\vert$$
$m$ and $n$ can take only integer values on $\soto$, and their physical
interpretation can be extracted from the asymptotic behavior at infinity,
leading to
$$m=\half (M+kN), \qquad n=\half (M-kN).\eqn\physintqn$$
Here $M$ and $N$ are the discrete momentum and winding number in the
asymptotically compact direction, respectively; both $N$ and $M$ are in
$\bf Z$.

The orbifold group action \tarorbkr\ corresponds in the parametrization
\suoneone\ to
$$\Theta:\quad r\rightarrow r, \qquad \theta_L\rightarrow \pi -\theta_R,
\qquad \theta_R \rightarrow \pi -\theta_L.\eqn\orbeuclactfund$$
Under the orbifold action, vertex operators with quantum numbers $m,n$ are
mapped to vertex operators with quantum numbers $-n,-m$, which clearly
corresponds to the expected behavior of the momentum and winding states under
the $\ztwo$.  The Jacobi functions $\CP^{\ell}_{mn}$ transform under
the $m\leftrightarrow -n$ interchange as follows [\vilen]:
$$\CP^{\ell}_{mn}(\cosh r)=\frac{\Gamma (\ell +n+1)
\Gamma (\ell -n+1)}{\Gamma (\ell +m+1) \Gamma (\ell
-m+1)}\CP^{\ell}_{-n,-m}(\cosh r),\eqn\jact$$
and we arrive at the following picture.  The orbifold group $\ztwo$ acts on
the untwisted sector, consisting of closed strings in the original black hole
background, by
$$\Theta :\quad\ket{T^{\ell}_{mn}}\;\rightarrow\;(-1)^{m+n}
\frac{\Gamma (\ell +n+1)\Gamma (\ell -n+1)}{\Gamma (\ell +m+1)
\Gamma (\ell -m+1)}\ket{T^{\ell}_{-n,-m}};
\eqn\ketaction$$
consequently, the only surviving tachyons are the following $\ztwo$-even
linear combinations of the tachyons with sharp momentum and winding number,
$$\frac{\Gamma (\ell +m+1)}{\Gamma (\ell +n+1)}\ket{T^{\ell}_{mn}}+
(-1)^{m+n}\frac{\Gamma (\ell -n+1)}{\Gamma (\ell -m+1)}
\ket{T^{\ell}_{-n,-m}}.\eqn\invclosedtach$$
Recalling the indications obtained within $c=1$ matrix models that amplitudes
simplify considerably when expressed in terms of renormalized vertex
operators, it may be natural to redefine e.g.\
$$T^{\ell}_{mn}(r,\theta_L,\theta_R)\rightarrow \CT ^{\ell}_{mn}\equiv
\frac{\Gamma(\ell +m+1)}{\Gamma(\ell +n+1)}
T^{\ell}_{mn}(r,\theta_L,\theta_R).\eqn\renorm$$
On these renormalized vertex operators, \jact\ can be rewritten as
$$\Theta:\quad\CT ^{\ell}_{mn}(r,\theta_L,\theta_R) \;\rightarrow\;
(-1)^{m+n}\CT ^{\ell}_{-n,-m}(r,\theta_L,\theta_R).\eqn\renact$$
The surviving primaries of the untwisted sector are thus given by
\invclosedtach .  To obtain the string spectrum, one has to impose the
equivariant BRST conditions \cohomolo .  Imposing the on-shell condition, we
get [\nelson] $m=\pm n =\pm (3\ell +\frac{3}{2})$.  This completes our
analysis of the closed-string tachyons.  The orbifold group action can be
extended easily to the discrete modes of [\nelson], as we know the action on
the primaries as well as on the Kac-Moody currents.

The twisted sector of our orbifold consists of open strings.  The orbifold
group action on the target fixes the boundary conditions on the worldsheet,
and determines the spectrum.  Primary states of the twisted sector are built
up from the representation theory of $\sltwo$; the left and right Kac-Moody
symmetries are now identified.  As $\theta_L=\pi -\theta_R$ at the boundary,
the open string will have only winding modes, moving in $\tilde{\theta}\equiv
\half (\theta_L-\theta_R)$:
$$\tilde{T}^{\ell}_{m}(r, \tilde{\theta})=\bra{\ell,m}g(r, \sfrac{\pi}{2}+
\tilde\theta ,\sfrac{\pi}{2}-\tilde\theta )\ket{\ell,-m}\eqn\optach$$
This is not unusual, an open string model with only winding modes in a flat
toroidal target has been constructed in [\wso,\dual].  In the coordinate
system $r,\theta$, the open-string modes \optach\ are effectively
one-dimensional, since they are restricted to
$$\theta =\pm\pi /2, \qquad r\ {\rm arbitrary}.\eqn\restropen$$
Far away from the black hole, the free sign in \restropen\ becomes a new
discrete quantum number of the open string, which is not conserved in
interactions with the black hole.  Analogously as in the untwisted sector,
we have to project to $\ztwo$ invariant states, and impose the on-shell
condition $m=\pm (3\ell +\frac{3}{2})$.  Analyzing the equivariant BRST
cohomology in the open sector, we also obtain an infinite tower of discrete
states restricted effectively to \restropen .

In worldsheet orbifold models, much information is gathered in the boundary
and crosscap states [\boundcross,\wso,\csworb], which allow one e.g.\ to find
the BRST invariant Chan-Paton factors of the open spectrum, hence determining
the Yang-Mills gauge symmetry.  Equations of motion for the open-string
background are determined from the condition of conformal invariance of the
boundary state,
$$(L_n-\bar{L}_{-n})|B\rangle =0,\qquad n\in {\bf Z}.\eqn\vir$$
Setting $n=0$ in \vir , we get a simple equation for the Kac-Moody primary
piece of the boundary operator:
$$\left( \frac{\p ^2}{\p \theta _L^2}-
\frac{\p ^2}{\p \theta _R^2}\right) |B\rangle_0=0.\eqn\virzero$$
Consequently, the primary part of the boundary operator can be written as a
sum of the contributions from the purely winding modes and purely momentum
modes,
$$|B\rangle_0=|\CB (\theta ,r)\rangle +| \widetilde{\CB}
(\widetilde{\theta},r)\rangle .\eqn\boun$$
We cannot, however, construct the full boundary and crosscap states and find
the proper Chan-Paton symmetry group of the black hole coset, as this issue
requires detailed knowledge of the closed string model, which is not yet
available.

If we trust the procedure of inferring the temperature of the black hole from
its Euclidean continuation by measuring the periodicity of $\theta$ far away
from the black hole, our orbifold truncation of the black hole cannot
change its temperature, which is thus the same as in the closed string coset.
On the other hand, what we do expect is a change in the temperature behavior
of the system in the vicinity of the Hagedorn temperature.  Indeed, there is a
serious difference between the formal high-temperature behavior of open and
closed string theories even in flat spacetime: The nonzero contribution of
the first winding mode to the boundary state is not compensated for by an
analogous contribution to the crosscap state, and causes BRST anomalies and
infinities already at the top of the Hagedorn temperature, where the closed
string model [\atick] is still formally finite.

Now let us turn to the Minkowski signature.  The orbifold truncates the
Minkowski black hole at $u=v$, where a time-like orbifold singularity is
formed.  Open strings of the model are confined to the orbifold singularity.
In the Euclidedan case we have learnt that the open-string tachyon has only
winding modes.  As there are no winding modes allowed in the Minkowski
signature, the whole spectrum of open strings will consist of discrete
states, which can be constructed precisely as in [\nelson].  Since the open
strings are confined with their center of mass to the orbifold singularity,
their corresponding background fields are restricted to the singularity as
well.  Hence, the effective low-energy Lagrangian is expected to acquire the
following form:
$$\CL_{\rm eff}=\int _Mdu\, dv\; e^{\Phi}\sqrt{G}\{ R+(\nabla \Phi )^2
+(\nabla T)^2-2T^2-8\} +\int_{\p M}d\tau \; e^{\Phi /2}\sqrt{G_{\tau\tau}}
\{ K+\ldots \}.\eqn\effact$$
We have denoted by $\p M$ the orbifold singularity in the truncated black
hole, $K$ denotes the exterior curvature of $\p M$ in $M$, and ``$\ldots$''
denote contributions from the open-string background fields.  In higher
dimensions, the boundary terms in the effective action would turn the
orbifold singularity to a dynamical membrane [\dlp,\membr].  In two
dimensions, the (massive) vector boson is one of the discrete states of the
open sector, carries the adjoint representation of the Chan-Paton symmetry
group, and its background plays the role of the Yang-Mills gauge field
(sitting at the orbifold singularity).  Hence, it is natural to expect that a
class of deformations of the truncated black hole will exist, related to
charged black holes (compare [\ucsb,\openbh]).  Note that in the dual
spacetime, the background gauge field is no longer restricted to the orbifold
singularity, as a result of general properties of open-string target duality
[\dual,\dlp] (see also our discussion of duality below).

The unitary gauge choice we have used up to now ceases to be valid in the
vicinity of the singularity at $uv=1$.  Witten has shown [\witten] that
in that region, the black hole is effectively described by a topological
field theory, namely the $U(1)$ Chern-Simons-Witten (CSW) theory,
dimensionally reduced from 3D to 2D.  Indeed, upon parametrizing the
spacetime manifold in the vicinity of $uv=1$ by $u=e^w$ and $v=e^{-w}$, the
gauged WZW Lagrangian reduces to
$$\CL_{\rm sing}=-\frac{k}{4\pi}\int d^2z\;\sqrt{h}\, h^{ij}\, D_ia\, D_jb+
\frac{ik}{4\pi}\int d^2z\; \sqrt{h}\, wF,\eqn\toposing$$
where $F=\epsilon^{ij}(\p _iA_j-\p _jA_i)$, $D_i$ is the covariant derivative
for $A$, and $a,b$ carry $U(1)$ charge $\pm 1$ respectively.  The second term
in \toposing\ is obviously the dimensional reduction of an abelian CSW
theory, $w$ being the third component of the 3D gauge field.  On our
truncated black hole, $w$ gets mapped by the orbifold group to minus itself.
Recalling that $A_z$ is mapped to $A_{\bar{z}}$, we easily find  the full
action of the orbifold group on the CSW theory, which fits nicely to the
picture already known in CSW theory on 3D $\ztwo$ orbifolds [\csworb].
Hence, we conclude that the vicinity of the singularity in the truncated
black hole is effectively described by a dimensionally reduced,
$\ztwo$-equivariant CSW theory.  The fact that topological degrees of freedom
become important in the vicinity of the spacetime singularity, i.e.\ in the
regime of the strong coupling, is indeed very interesting conceptually
[\eguchi].  In our case, however, the correspondence just discussed may have
an even more direct impact, as CSW theory on $\ztwo$ orbifolds is helpful in
clarifying such issues of open string theory as the structure of the
non-anomalous Chan-Paton factors for the model, relating them to 3D algebraic
topology [\csworb].

The truncated black hole model inherits a spacetime duality symmetry from its
closed string ancestor.  Here we will establish a simple, yet striking result,
that the truncated black hole is dual to a conventional open string theory
in the black hole background.  So far we have considered the model in the
axial gauging representation, in which the $\ztwo$ orbifold group acts by $g
\rightarrow sg^{-1}s$ (see \actwso ).  In the dual model, the only difference
is in the action of the $U(1)$ gauge group, which is now
$$g\rightarrow h\, g\, h^{-1}, \eqn\dualact$$
while the $\ztwo$ group action remains unchanged.  While in the truncated
black hole model the $\ztwo$ orbifold action preserves the generator of the
$\sooo$ gauge symmetry, and extends thus the gauge group to $\ztwo\times
\sooo$, in the dual model the discrete group action reverses the generator of
the gauge group.  Consequently, the dual geometry is described by a gauged
WZW model, in which the abelian gauge group is extended to the semi-direct
product $\ztwo\semi\sooo$.

We can get some insight into the dual geometry by switching to its
alternative description in the axial gauging representation.  The
$O(1,1)\equiv\ztwo\semi\sooo$ gauge group now acts by
$$SO(2):\quad \pmatrix{a&u\cr-v&b\cr}\rightarrow
h\pmatrix{a&u\cr-v&b\cr}h,\qquad \ztwo :\quad \pmatrix{a&u\cr-v&b\cr}
\rightarrow \pmatrix{b&u\cr -v&a\cr}.\eqn\dualaxial$$
Recalling the gauge fixing conditions discussed above, we can see that the
$\ztwo$ acts on the Minkowski black hole (in the Kruskal coordinate system)
by
$$(u,v)\rightarrow\left\{\eqalign{(u,v),\qquad\ \ \ \ \ &{\rm regions\
I-IV},\cr (-u,-v),\qquad &{\rm regions\ V,VI}.\cr}\right. \eqn\dualactkrusk$$
In Euclidean signature the situation is analogous. The truncated semi-infinite
cigar, on which the orbifold group acts by $(r,\theta )\rightarrow (r,\pi -
\theta )$, is dual to the infinite funnel, on which the target part of the
orbifold group acts trivially.

A few remarks on the dual description of the truncated black hole are in
order.  First, the $\ztwo$ group action of \dualaxial\ leaves points of
regions I$-$IV fixed, and acts only on the worldsheet -- the model describes
in these regions the conventional open string theory in the black hole
geometry.  Indeed, the open-string winding modes, which are fixed to $\theta
=\pm\pi /2$ by \restropen , become momentum modes in the dual picture, and
can move freely in the dual coordinate $\tilde{\theta}$.  Second, it is
natural that while in the truncated black hole we gauge the direct product
$\ztwo\times\sooo$, in the dual model we have to gauge the semi-direct
product $\ztwo\semi\sooo$.  Indeed, the duality transformation exchanges the
generator of the gauge group action on the group manifold with the Killing
vector of the model, and the $\ztwo$ action of worldsheet orbifold models
changes the orientation of precisely one of these two vector fields, as can
be seen e.g.\ in the semiclassical limit.  Finally, note that this pattern of
dual models can be extended easily along the lines of [\rocek] to any target
for open string propagation interpretable as a sigma model with a Killing
vector.

Let us now turn to possible relations of our truncated 2D black hole to $c=1$
string theory.  It was realized in the free-fermion approach to $c=1$ string
theory [\polch] that the eigenvalue density fluctuations (i.e.\ the spacetime
tachyon) can escape into the world beyond the mirror, once exposed to not
very extreme conditions.  This was related in [\marsh] to a possibility for
the string in the black hole background to fall into the black hole, be
scattered off the singularity, and enter the second asymptotically flat
domain in the direction of the increasing Schwarzschild time.  In the
free-fermion framework one can start from the free fermionic system in the
upside-down harmonic oscillator potential, with infinite walls at distances
$\pm A$ before the double-scaling limit.  There is a simple variation of this
theory [\moore] in which one fixes an alternative set of boundary conditions
on the potential by moving one of the two walls (say) to the origin.  We
believe that such a modification of the standard double-scaled theory of free
fermions can be related to our truncated black hole.  If this conjecture is
true, one can possibly learn something about the wall of $c=1$ string theory
from the fact that in the analogous situation of the black hole geometry, the
wall is inhabited by open strings.  This might be a challenge for the matrix
model approach to $c=1$ string theory.

In this paper we have found a class of exact black-hole solutions of two
dimensional (closed) string theory, with open strings in the spectrum.
One lesson we can draw from our results is a reconfirmation of the general
observation that in string theory, the worldsheet and target geometries are
tightly entangled with each other.  Indeed, asking for a black hole geometry
with one asymptotically flat domain, we end up with a theory of open and
closed unoriented strings, i.e.\ we are in some sense forced to change the
topology of the string.  In this paper we have limited ourselves to
establishing this result, and analyzed only briefly the physics and geometry
of the truncated black hole.  Since many aspects of the closed-string black
hole itself, as well as its precise relation to $c=1$ matrix models are still
unclear, many of the exciting questions one can ask about the role of the
truncated black hole in two dimensional string theory cannot be answered
until we gain some more insight into these issues, and represent an
interesting task for the future.

{\bf Acknowledgement.}  It is a pleasure to acknowledge stimulating
discussions with E. Martinec, S. Shenker and G. 't Hooft at an early stage of
this work.  The author is grateful to the organizers of the NATO ASI on ``New
Symmetry Principles in Quantum Field Theories'' (July 1991) for their
hospitality in Carg\`{e}se, where part of this work was done.
%
%
\refout
\end
\bye